\documentclass[showpacs,showkeys,twocolumn]{revtex4}

\usepackage{amsfonts,amsmath,amssymb,bm,hyperref,graphicx,dsfont}

\begin{document}


\title{Neutrino spin-flavor oscillations in rapidly
varying external fields}

\author{Maxim Dvornikov}
\affiliation{Institute of Terrestrial Magnetism, Ionosphere and
\\ Radiowave Propagation (IZMIRAN) \\ 142190, Troitsk, Moscow
region, Russia} \email{maxdvo@izmiran.ru}

\date{\today}

\begin{abstract}
We study neutrino spin-flavor oscillations in presence of rapidly
varying external fields. The general formalism for the description
of neutrino oscillations in arbitrary rapidly varying external
fields is elaborated. We obtain the effective Hamiltonian which
determines the evolution of the averaged neutrino wave function.
We apply the general technique to neutrino oscillations in rapidly
varying magnetic fields. The special case of the constant
transversal and twisting magnetic fields is studied. We evaluate
the effect of neutrino spin-flavor oscillations in rapidly varying
magnetic fields of the Sun. The numerical solutions of the
Schr\"{o}dinger equation with the Hamiltonian accounting for the
constant transversal and twisting magnetic fields are presented.
We compare them with our approximate analytical transition
probability formula and reveal good agreement at high frequencies
of the twisting magnetic field.
\end{abstract}



\maketitle

\section{Introduction}

For the first time the idea of neutrino oscillations was
theoretically predicted in Ref.~\cite{Pon58eng}. Since then many
experimental and theoretical studies of neutrino oscillations have
been carried out. One of the most interesting problems in neutrino
physics is the solar neutrino deficit. Nowadays it is
experimentally established (see, e.g., Ref.~\cite{Ahm04}) that the
disappearance of solar electron neutrinos can be accounted for by
the LMA-MSW solution \cite{Wol78,MikSmi85eng}. The process of
neutrino oscillations is likely to be the most reliable
explanation of the solar neutrino problem. There are several
theoretical models of solar neutrino oscillations such as
mentioned above LMA-MSW solution. However other scenarios like
spin-flavor precession (see
Refs.~\cite{LimMar88,Akh88PL,Smi91,AkhPetSmi93,LikStu95JETPeng})
are also considered.

Recently resonant neutrino spin-flavor oscillations were studied
in Ref.~\cite{ChaPul04} where an attempt was made to reproduce the
data of the major solar neutrino experiments using the peak
profiles of the solar magnetic field. Different approach to the
solar neutrino problem was made in Ref.~\cite{AkhPul03}. Supposing
that neutrino spin-flavor precession played a subdominant role in
comparison with flavor oscillations the combined action of these
two mechanisms was examined. It was also possible to impose the
constraints on the characteristics of a neutrino, namely to
restrict its magnetic moment, and the strength of the solar
magnetic field. Solar electron neutrino transitions into a
non-electron antineutrino were studied in Ref.~\cite{KanKim04}. On
the basis of the resent SNO data the indication of the neutrino
Majorana magnetic moment was obtained in that paper. A global
analysis of the solar neutrino problem solution with help of the
spin-flavor precession was given in Ref.~\cite{BarMirRasSemVal02}.
Two flavor oscillations scenario in the optimized self-consistent
magneto-hydrodynamical magnetic field profile was adopted. A
careful analysis of the predicted solar neutrino fluxes was
performed in Ref.~\cite{CouTurKos02} using up to date
helioseismological data.

We studied neutrino oscillations in electromagnetic fields of
various configurations in our previous works. First it is
necessary to mention that the Lorentz invariant formalism for the
description of neutrino spin-flavor oscillations was elaborated in
Ref.~\cite{EgoLobStu00}. It allows one to study neutrino
oscillations in arbitrary electromagnetic fields. For example,
neutrino oscillations in an electromagnetic wave and in a
longitudinal, with respect to the neutrino momentum, magnetic
field were considered in Ref.~\cite{EgoLobStu00}. In
Ref.~\cite{DvoStu01YFeng} we worked out the method for the
investigation of the neutrino evolution equation solution near the
resonance point. This technique is of great importance when the
evolution equation cannot be solved analytically. In
Ref.~\cite{DvoStu04YFeng} we have applied this method to the
studying of the parametric resonance in neutrino oscillations in
amplitude modulated electromagnetic wave. The major feature of
Refs.~\cite{DvoStu01YFeng,DvoStu04YFeng} was the use of the
perturbation theory in solving the neutrino evolution equation. It
should be noted that the perturbation theory was also used in
Ref.~\cite{FisGas01} where the parametric resonance in neutrino
oscillations in matter with periodically varying density was
studied. In Ref.~\cite{DvoStu02JHEP} we proposed the Lorentz
invariant quasi-classical approach for the description of neutrino
spin oscillations in arbitrary non-derivative external fields.
This method was applied in Ref.~\cite{DvoGriStu05} to the
description of neutrino spin oscillations in gravitational fields.

In this paper we study neutrino oscillations in presence of
general rapidly varying fields. It should be noted that the
influence of rapidly varying fields on mechanical oscillations was
investigated in Ref.~\cite{Kap51}. It is also well known that
there are certain analogies between mechanical and neutrino
oscillations (see, for instance, Ref.~\cite{Wei87}). In
Sec.~\ref{GF} we start from the neutrino evolution equation with
the Hamiltonian which accounts for the neutrino interaction with
rapidly varying fields. Note that we do not fix the explicit form
of the Hamiltonian. Then we derive the new effective Hamiltonian
governing the time evolution of the averaged neutrino wave
function. Therefore the obtained new Hamiltonian allows one to
study the neutrino conversion in presence of arbitrary rapidly
varying external fields. Our result is also beyond the
perturbation theory because no assumptions about the smallness of
rapidly varying fields are made. Then, in Sec.~\ref{NSFOB}, we
apply the general technique to neutrino oscillations in rapidly
varying magnetic fields. We investigate the combination of
constant transversal and twisting magnetic fields. The neutrino
conversion in solar magnetic fields is discussed. In Sec.~\ref{NS}
we present the numerical solutions of the Schr\"{o}dinger equation
for the neutrino system interacting with constant transversal and
twisting magnetic fields and compare them with the approximate
analytical solutions found in the present work.

\section{General formalism}\label{GF}

Let us consider the evolution of the two neutrinos $\nu=(\nu_1,
\nu_2)$, which can belong to different flavors and helicity
states. The evolution of the system is described by the
Schr\"{o}dinger type differential equation,
\begin{equation}\label{shr}
  i\frac{\mathrm{d}\nu}{\mathrm{d}t}=H\nu,
\end{equation}
where the Hamiltonian $H$ involves external fields, e.g.,
electromagnetic field, interaction with matter and the
characteristics of the neutrinos. Here we do not specify the
explicit form of the Hamiltonian but just suppose that it is
decomposed into two terms,
\begin{equation}\label{ham}
  H=H_0+\mathcal{H},
  \quad
  \mathcal{H}(t+T)=\mathcal{H}(t).
\end{equation}
The first term in Eq.~\eqref{ham} -- $H_0$, corresponds to the
neutrino interaction with constant or slowly varying external
fields. In presence of only this term the solution of
Eq.~\eqref{shr} can be easily found. The solution is known to be
periodical with the typical frequency $\Omega_0\sim
1/L_\mathrm{eff}$, where $L_\mathrm{eff}$ is the oscillations
length. The second term in Eq.~\eqref{ham} -- $\mathcal{H}(t)$,
corresponds to rapidly varying external fields. The frequency
$\omega=2\pi/T$ should be much greater than $\Omega_0$:
$\omega\gg\Omega_0$. For example, we can suppose that the
components of the Hamiltonian depend on time like
$\mathcal{H}(t)\sim \cos\omega t$ or $\sin\omega t$. Note that we
do not make any assumptions about the strength of varying external
fields.

We will seek the solution of the Eqs.~\eqref{shr} and \eqref{ham}
in the form (see also Ref.~\cite{Kap51}),
\begin{equation}\label{formsol}
  \nu(t)=\nu_0(t)+\xi(t).
\end{equation}
In Eq.~\eqref{formsol} the function $\xi(t)$ is the small rapidly
oscillating one with zero mean value. The mean value $\bar{f}$ of
a function $f(t)$ is the time averaging over the period $T$. The
function $\nu_0(t)$ in Eq.~\eqref{formsol} is the slowly varying
one during $T$. Substituting Eq.~\eqref{formsol} in
Eq.~\eqref{shr} we obtain
\begin{equation}\label{medeq}
  i\dot{\nu}_0+i\dot{\xi}=H_0\nu_0+\mathcal{H}\nu_0+
  \mathcal{H}\xi+H_0\xi.
\end{equation}
There are two groups of the terms in Eq.~\eqref{medeq}: rapidly
and slowly varying ones. In order for this expression to be an
identity the terms of each group in the left-handed side of the
equation should be equal to the terms of the same group in the
right-handed side.

First let us discuss rapidly varying terms. The function
$\dot{\xi}$ standing in the left-handed side belongs to this
group. Despite the function $\xi$ is small, its derivative can be,
in principle, not small. Indeed, it is proportional to $\omega$
and thus $\dot{\xi}$ is the great quantity. There are three
rapidly varying terms in the right-handed side of
Eq.~\eqref{medeq}, namely $\mathcal{H}\nu_0$, $\mathcal{H}\xi$ and
$H_0\xi$. However the second and the third terms in this group are
the small ones because they contain the small factor $\xi$. Thus
we can drop the terms $\mathcal{H}\xi$ and $H_0\xi$ when we
discuss rapidly varying terms. Finally one receives the equation
for the function $\xi$
\begin{equation}\label{medeqxi}
  i\frac{\mathrm{d}\xi}{\mathrm{d}t}=\mathcal{H}\nu_0.
\end{equation}
It should be noted that the terms in Eq.~\eqref{medeq} containing
$\xi$ can contribute to the phase of the wave function if long
evolution times are considered. However, when we discuss possible
applications of this approach in Sec.~\ref{NSFOB}, only one period
of the transition probability variation is taken into account.
This problem is also analyzed in Sec.~\ref{NS}. Equation
\eqref{medeqxi} can be easily integrated and we find its solution
\begin{equation}\label{xisol}
  \xi(t)=-i
  \left(
  \int\mathcal{H}(t)\thinspace\mathrm{d}t
  \right)
  \nu_0(t).
\end{equation}
Here we take into account that variations of the function $\nu_0$
are small. Therefore Eq.~\eqref{xisol} presents the time
dependence of the function $\xi$.

Now let us proceed to the description of $\nu_0$ evolution. We can
omit $i\dot{\xi}$ and $\mathcal{H}\nu_0$ in Eq.~\eqref{medeq}
because we have already proved that these terms are equal [see
Eq.~\eqref{medeqxi}]. Averaging residual terms in
Eq.~\eqref{medeq} over the period $T$ and allowing for
\[
  \overline{\dot{\nu}_0}=\dot{{\nu}}_0,
  \quad
  \overline{H_0\nu_0}=H_0\nu_0,
  \quad
  \overline{H_0\xi}=H_0\overline{\xi}=0,
\]
we obtain the equation for the function $\nu_0$
\begin{equation}\label{nu0interm}
  i\frac{\mathrm{d}\nu_0}{\mathrm{d}t}=
  H_0\nu_0+\overline{\mathcal{H}\xi}.
\end{equation}
We cannot neglect the term $\overline{\mathcal{H}\xi}$ in
Eq.~\eqref{nu0interm} since it is the product of two periodically
varying functions. The mean value of this quantity is not equal to
zero. With help of Eq.~\eqref{xisol} equation \eqref{nu0interm}
can transformed into the more common form
\begin{equation}\label{eqnu0}
  i\frac{\mathrm{d}\nu_0}{\mathrm{d}t}=
  H_\mathrm{eff}\nu_0,
\end{equation}
where
\begin{equation}\label{Heff}
  H_\mathrm{eff}=H_0-i
  \overline{\mathcal{H}
  \left(
  \smallint\mathcal{H}\mathrm{d}t
  \right)}.
\end{equation}

In derivation of Eqs.~\eqref{eqnu0} and \eqref{Heff} we do not
make any assumptions about the smallness of the interaction
described by the Hamiltonian $\mathcal{H}$. Thus our result is
beyond the perturbation theory used in
Refs.~\cite{DvoStu01YFeng,DvoStu04YFeng,FisGas01}.

The Hamiltonian $H$ can be always represented in the form,
\begin{equation*}
  H=(\bm{\sigma}\cdot\mathbf{h}),
\end{equation*}
where $\mathbf{h}$ is the arbitrary $3D$-vector. Using this
representation we discuss two cases:
\begin{itemize}
  \item $H_0=(\bm{\sigma}_\bot\cdot\mathbf{h}_\bot)$,
  $\mathcal{H}=\sigma_3 h_3$,
  \item $\mathcal{H}=(\bm{\sigma}_\bot\cdot\mathbf{h}_\bot)$,
  $H_0=\sigma_3 h_3$.
\end{itemize}
Here we introduce $2D$-vectors
$\bm{\sigma}_\bot=(\sigma_1,\sigma_2)$ and
$\mathbf{h}_\bot=(h_1,h_2)$.

In the former case the application of the developed in the present
paper technique leads to no interesting consequences. Indeed,
using Eq.~\eqref{Heff} we find that
\begin{equation*}
  H_\mathrm{eff}=(\bm{\sigma}_\bot\cdot\mathbf{h}_\bot)-
  i\widehat{\mathds{1}}
  \overline{h_3
  \left(
  \smallint h_3\mathrm{d}t
  \right)},
\end{equation*}
where $\widehat{\mathds{1}}$ is the unit matrix. It is well known
that terms in the Hamiltonian proportional to the unit matrix can
be omitted. For example, if we study neutrino oscillations in
matter with rapidly varying density $n(t)$, it will cause no
essential effect on the oscillations process. However, if one
accounts for all components of the vector $\mathbf{h}$ in
$\mathcal{H}$, i.e. $\mathcal{H}=(\bm{\sigma}\cdot\mathbf{h})$, it
could result in some interesting effects. This case requires
special careful examination.

Now we consider the latter case. First we introduce two auxiliary
$2D$-vectors $\mathbf{a}=\mathbf{h}_\bot$ and
$\mathbf{b}=\smallint \mathbf{a}\thinspace\mathrm{d}t$. Again
using Eq.~\eqref{Heff} we obtain that
\begin{align*}
  H_\mathrm{eff}= & \sigma_3 h_3-
  i\widehat{\mathds{1}}
  \overline{(\mathbf{a}\cdot\mathbf{b})}+
  \sigma_3\overline{(a_1 b_2-a_2 b_1)}
  \\
  & \to
  \sigma_3
  \left[
  h_3+
  \overline{(a_1 b_2-a_2 b_1)}
  \right].
\end{align*}
Therefore the off-diagonal rapidly varying terms can shift the
resonance point. It was shown in Ref.~\cite{DvoStu02JHEP} that
both axial-vector and tensor and pseudotensor interactions can
cause the neutrino spin precession. Thus rapidly varying external
fields of the mentioned above types, e.g., electromagnetic fields,
interaction with moving and polarized matter etc., will
nontrivially affect the neutrino oscillation process. We study
below one the possible examples.

\section{Neutrino spin-flavor oscillations in
rapidly varying magnetic fields}\label{NSFOB}

In this section let us discuss the neutrino evolution in matter
under the influence of a combination of the two types of magnetic
fields
\begin{itemize}
  \item Constant transversal magnetic field $\mathbf{B}_0$,
  \item Twisting magnetic field $\mathbf{B}(\mathbf{r})$.
\end{itemize}
Note that if a magnetic field is constant in time we may consider
only its transversal component with respect to the neutrino
velocity. These magnetic field configurations were studied
separately in the majority of works devoted to neutrino
spin-flavor oscillations. For instance, only constant transversal
component of a magnetic field was taken into account in
Refs.~\cite{LimMar88,Akh88PL}, and the effect of only twisting
magnetic field on neutrino oscillations was considered in
Refs.~\cite{Smi91,AkhPetSmi93,LikStu95JETPeng}.

In our case the Hamiltonians $H_0$ and $\mathcal{H}$ have the form
\begin{equation}\label{B0ham}
  H_0=
  \begin{pmatrix}
    V/2 & \mu B_0 \\
    \mu B_0 & -V/2 \
  \end{pmatrix},
\end{equation}
and
\begin{equation}\label{Bham}
  \mathcal{H}=
  \begin{pmatrix}
    0 & \mu B e^{-i\omega t} \\
    \mu B e^{i\omega t} & 0 \
  \end{pmatrix}.
\end{equation}
where $\mu$ is the neutrino magnetic moment. In Eq.~\eqref{B0ham}
we introduced the quantity,
\begin{equation*}
  \frac{V}{2}=
  \frac{\Delta m^2}{4E}\Theta-
  \frac{G_F}{\sqrt{2}}
  n_\mathrm{eff},
\end{equation*}
where $\Theta$ is the function of the vacuum mixing angle
$\theta_\mathrm{vac}$ (the explicit form of $\Theta$ for various
transitions of the $\nu_{iL}\leftrightarrow\nu_{jR}$ type can be
found in Ref.~\cite{LikStu95JETPeng}), $\Delta m^2$ is the
difference of the neutrino mass squared, $E$ is the neutrino
energy, $n_\mathrm{eff}$ is the effective matter density, $\omega$
is the frequency of the transversal magnetic field variation,
$G_F$ is the Fermi constant.

Using Eqs.~\eqref{Heff}-\eqref{Bham} it is possible to derive the
expression for $H_\mathrm{eff}$,
\begin{equation}\label{Bhameff}
  H_\mathrm{eff}=
  \begin{pmatrix}
    V/2-(\mu B)^2/\omega & \mu B_0 \\
    \mu B_0 & -V/2+(\mu B)^2/\omega \
  \end{pmatrix}.
\end{equation}
It should be noted that the parameters $B$ and $B_0$ must have non
zero values. However it is interesting to discuss the case $B_0\to
0$. When a neutrino interacts only with twisting magnetic field
the problem of neutrino oscillations can be solved exactly (see,
e.g., Ref.~\cite{Smi91}). One can explicitly write the formula for
the transition probability
\begin{equation*}
  P(t)=
  \left(
    \frac{\mu B}{\Omega'}
  \right)^2
  \sin^2\Omega' t,
\end{equation*}
where $\Omega'= \sqrt{(V - \omega)^2/4 + (\mu B)^2}$. It can be
shown that in the limit of rapidly varying twisting magnetic field
the transition probability approaches to zero. Indeed
\begin{equation*}
  \left.
    P(t)
  \right|_{\omega\gg \mu B, V}\to
  \left.
    \left(
      \frac{2\mu B}{\omega}
    \right)^2
    \sin^2
      \left(
        \frac{\omega}{2} t
      \right)
  \right|_{\omega\gg \mu B, V}\to 0.
\end{equation*}
Thus we can see that the transition probability vanishes when a
neutrino interacts only with rapidly varying twisting magnetic
field. This result also follows from Eq.~\eqref{Bhameff}.

Since $B_0$ and $B$ do not depend on time in Eq.~\eqref{Bhameff},
we can solve Eq.~\eqref{eqnu0} for the case of the effective
Hamiltonian given in Eq.~\eqref{Bhameff}. The transition
probability is expressed in the following way,
\begin{equation}\label{transprob}
  P(t)=A\sin^2
  \left(
  \frac{\pi t}{L}
  \right),
\end{equation}
where
\begin{equation}\label{A}
  A=\frac{(\mu B_0)^2}
  {
  \left[
  V/2-(\mu B)^2/\omega
  \right]^2+
  (\mu B_0)^2},
\end{equation}
and
\begin{equation}\label{osclength}
  \frac{\pi}{L}=\Omega=
  \sqrt{
  \left[
  V/2-(\mu B)^2/\omega
  \right]^2+
  (\mu B_0)^2}.
\end{equation}

Thus the validity of the developed method (the frequency of the
twisting magnetic field $\omega$ should be much greater than the
characteristic "frequency" of the system at the absence of this
field) can be expressed as the constraint,
\begin{equation*}
  \omega\gg\Omega_0=
  \sqrt{
  (V/2)^2+
  (\mu B_0)^2}.
\end{equation*}

Now let us discuss the resonance conditions in neutrino
oscillations in rapidly varying magnetic fields. From
Eqs.~\eqref{transprob} and \eqref{A} it follows that if the
condition is satisfied,
\begin{equation}\label{res}
  \frac{V}{2}\simeq\frac{(\mu B)^2}{\omega},
\end{equation}
then $A\simeq 1$ and the transition probability can achieve great
values. This phenomenon is analogous to resonance amplification of
spin-flavor oscillations. However, in our case the resonance is
attained not due to the zero value of the parameter $V$.

We again emphasize that the strength of the twisting magnetic
field $B$ can be not small in the proposed technique. The only
assumption made consists in the great value of $\omega$. Therefore
we can discuss two cases:
\begin{itemize}
  \item $B_0\gg B$,
  \item $B\gg B_0$.
\end{itemize}
In the former case the value of $A$ in Eq.~\eqref{A} is great
independently of $B$. This case corresponds to the usual
spin-flavor precession in constant transversal magnetic field.
However, in the latter case the value of $A_0=A(B=0)$ is much less
than unity. Hence the transition probability at the absence of the
additional twisting magnetic field is small. In this case we can
choose the parameters so [see Eq.~\eqref{res}] that the amplitude
of the transition probability is great. It is this situation which
is of interest because the essential neutrino spin-flavor
conversion is attained under the influence of compact
perturbations of a magnetic field against a rather weak constant
background magnetic field. Such magnetic field configurations can
be created experimentally or observed in some astrophysical media.
Magnetic field in perturbations can have twisting structure. From
Eqs.~\eqref{A}-\eqref{res} one can derive the restriction imposed
on the $B$ and $B_0$,
\begin{equation}\label{BB0restr}
  \left(
  \frac{B}{B_0}
  \right)^2
  \gg 1.
\end{equation}

\begin{figure}
  \centering
  \includegraphics[scale=.7]{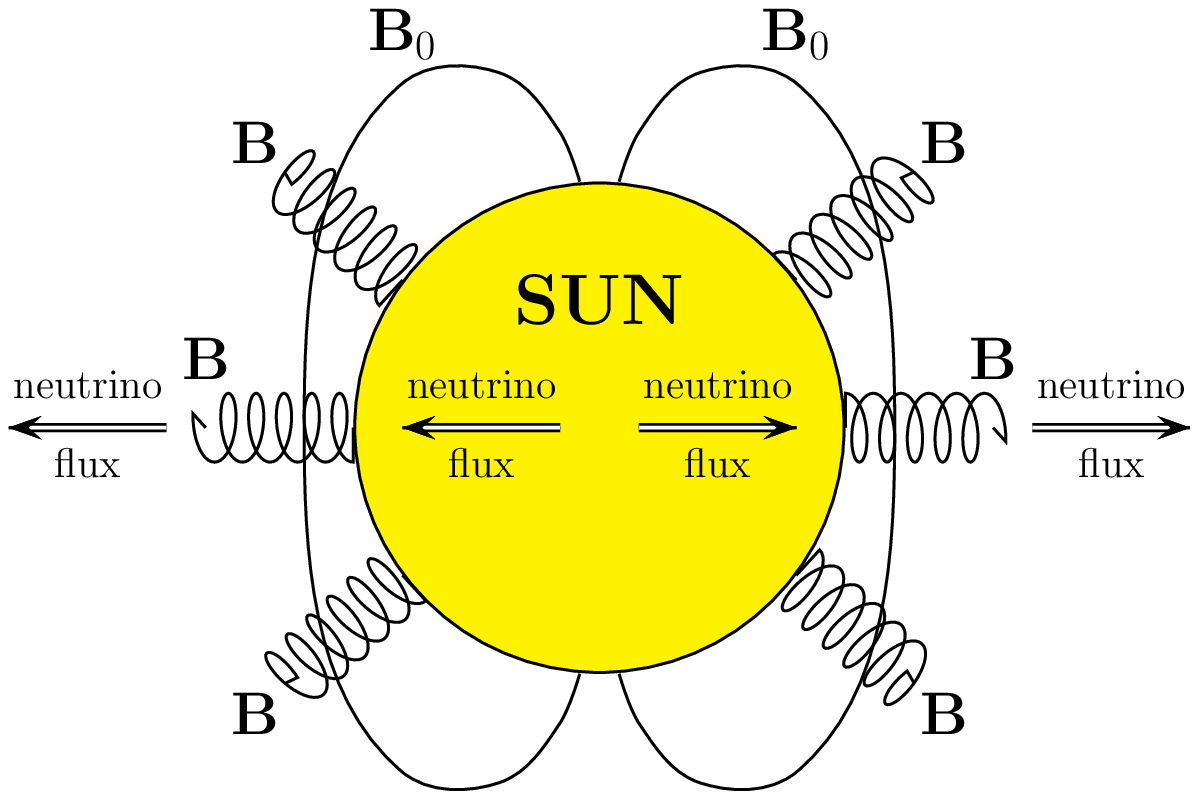}
  \caption{Neutrino flux propagation in the magnetic fields of the Sun,
  $\mathbf{B}_0$ and $\mathbf{B}$ are the vectors of constant tranversal
  and twisting magnetic fields.}
  \label{sunfig}
\end{figure}

Now we consider the possible application of the developed
technique to spin-flavor neutrino oscillations in magnetic fields
of the Sun. The solar magnetic field is unlikely to be only either
constant transversal or twisting. Therefore we can apply the
method elaborated in this paper to the description of the solar
neutrino conversion. The propagation of the neutrino flux in the
combination of constant transversal and twisting solar magnetic
fields is schematically depicted in Fig.~\ref{sunfig}.
We consider one of the possible channels of neutrino oscillations,
namely $\nu_{e L}\leftrightarrow\nu_{\mu R}$ conversion. First we
should estimate the parameter $V$ in Eq.~\eqref{B0ham}. In our
case the function $\Theta$ and the effective matter density have
the form (see, e.g., Ref.~\cite{LikStu95JETPeng}),
\begin{equation*}
  \Theta=
  \frac{1+\cos 2\theta_\mathrm{vac}}{2},
  \quad
  n_\mathrm{eff}=
  \left(
  n_e-\frac{1}{2}n_n
  \right).
\end{equation*}
Let us discuss a neutrino with the following properties: $\Delta
m^2\sim 10^{-5}\thinspace\text{eV}^2$, $\theta_\mathrm{vac}\sim
\pi/4$, which are not excluded by the modern experimental results
(see, for instance, Ref.~\cite{Eid04}). We take the neutrino
energy $E\sim 10\thinspace\text{MeV}$, which corresponds to the
$^{8}\text{B}$ solar neutrinos. Matter is supposed to consist
mainly of the hydrogen, i.e. $n_n\simeq 0$, $n_e=n_p$ (as a
consequence of the system electroneutrality) and has the density
$d\sim 1.4\thinspace\text{g}\cdot\text{cm}^{-3}$, which is close
to the mean density of the solar matter. For these parameters we
obtain that $V/2\sim 10^{-15}\thinspace\text{eV}$.

Now let us evaluate the strength of the magnetic fields, $B_0$ and
$B$, necessary for the $10\%$ conversion of the initial $\nu_{e
L}$ beam. We suppose that neutrinos conversion occurs along the
distance $D=t\simeq 9.2\times 10^{-2}R_\odot\simeq 3.2\times
10^{14}\thinspace\text{eV}^{-1}$, where $R_\odot$ is the solar
radius. We also suppose that neutrinos have the transitional
magnetic moment $\mu=10^{-11}\mu_\mathrm{B}$, where
$\mu_\mathrm{B}$ is the Bohr magneton. Setting $P(t)=0.1$ in
Eq.~\eqref{transprob} we obtain $B_0\simeq
17.6\thinspace\text{kG}$. From the condition~\eqref{BB0restr} we
can derive the strength of the twisting magnetic field: $B\simeq
3.2 B_0\simeq 56.3\thinspace\text{kG}$. The resonance
condition~\eqref{res} is satisfied if $\omega\simeq 1.0\times
10^{-14}\thinspace\text{eV}$. The close values of the twisting
magnetic field frequencies ($3\times 10^{-15}\thinspace\text{eV}$)
were considered in Ref.~\cite{AkhPetSmi93}.

Background matter perturbations can cause twisting magnetic fields
generation. If a neutrino interacts with solar plasma, the
frequencies of these magnetic fields can be evaluated with help of
the magnetic hydrodynamics (MHD) methods. We suppose that the
twisting magnetic field is induced by a MHD wave with the wave
vector perpendicular to $\mathbf{B}_0$. The frequencies of such
MHD waves can achieve values up to $1.5\times
10^5\thinspace\text{eV}$  for $B_0\simeq 17.6\thinspace\text{kG}$
(see, for instance, Ref.~\cite{AleBogRuk78p105eng}). The frequency
of the twisting magnetic field, used in the present work,
satisfies this condition. It should be noticed that solar plasma
fluctuations were studied in Ref.~\cite{BurDzhRasSemVal04}.

\section{Numerical simulations}\label{NS}

The analysis carried out in the previous section shows that the
influence of rapidly varying magnetic fields on neutrino
oscillations is very important in various astrophysical
environments, for example, in $\nu_{e L}\leftrightarrow\nu_{\mu
R}$ oscillations in magnetic fields of the Sun. In order to
substantiate the correctness of the approach developed in this
paper for the description of this problem we obtain the numerical
solutions of Eq.~\eqref{shr} with the Hamiltonian given in
Eqs.~\eqref{B0ham} and \eqref{Bham}. The numerical solutions
account for all terms in Eq.~\eqref{medeq}. Then we get the
transition probabilities on the basis of the numerical solutions.
We also compare these numerical transition probabilities with our
approximate formula for the transition probability
[Eq.~\eqref{transprob}] which is derived in the limit of rapidly
varying Hamiltonian $\mathcal{H}$. The numerical simulation is
carried out with MATLAB language for technical computing (version
6.5).

\begin{figure}
  \centering
  \includegraphics[scale=.45]{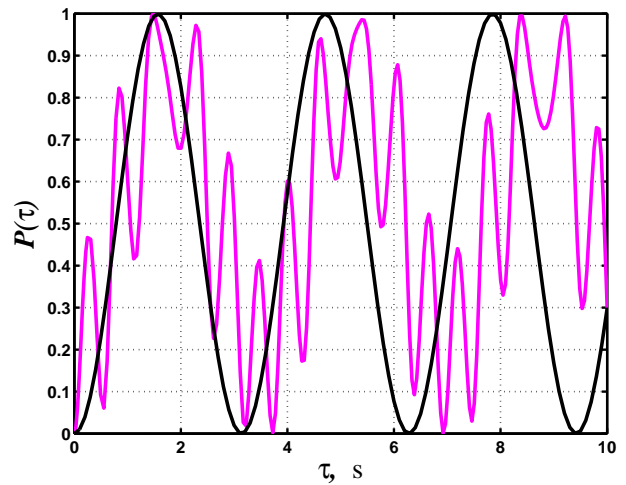}
  \caption{Neutrino transition probability for
  $\nu_{e L}\leftrightarrow\nu_{\mu R}$ oscillations,
  $\omega=1.0\times10^{-14}\thinspace\text{eV}$. The magenta line
  corresponds to the transition probability found with help of
  the numerical solution of Eq.~\eqref{shr}, the black line
  represents the transition probability
  given in Eq.~\eqref{transprob}.}
  \label{numtrpr1}
\end{figure}

\begin{figure}
  \centering
  \includegraphics[scale=.45]{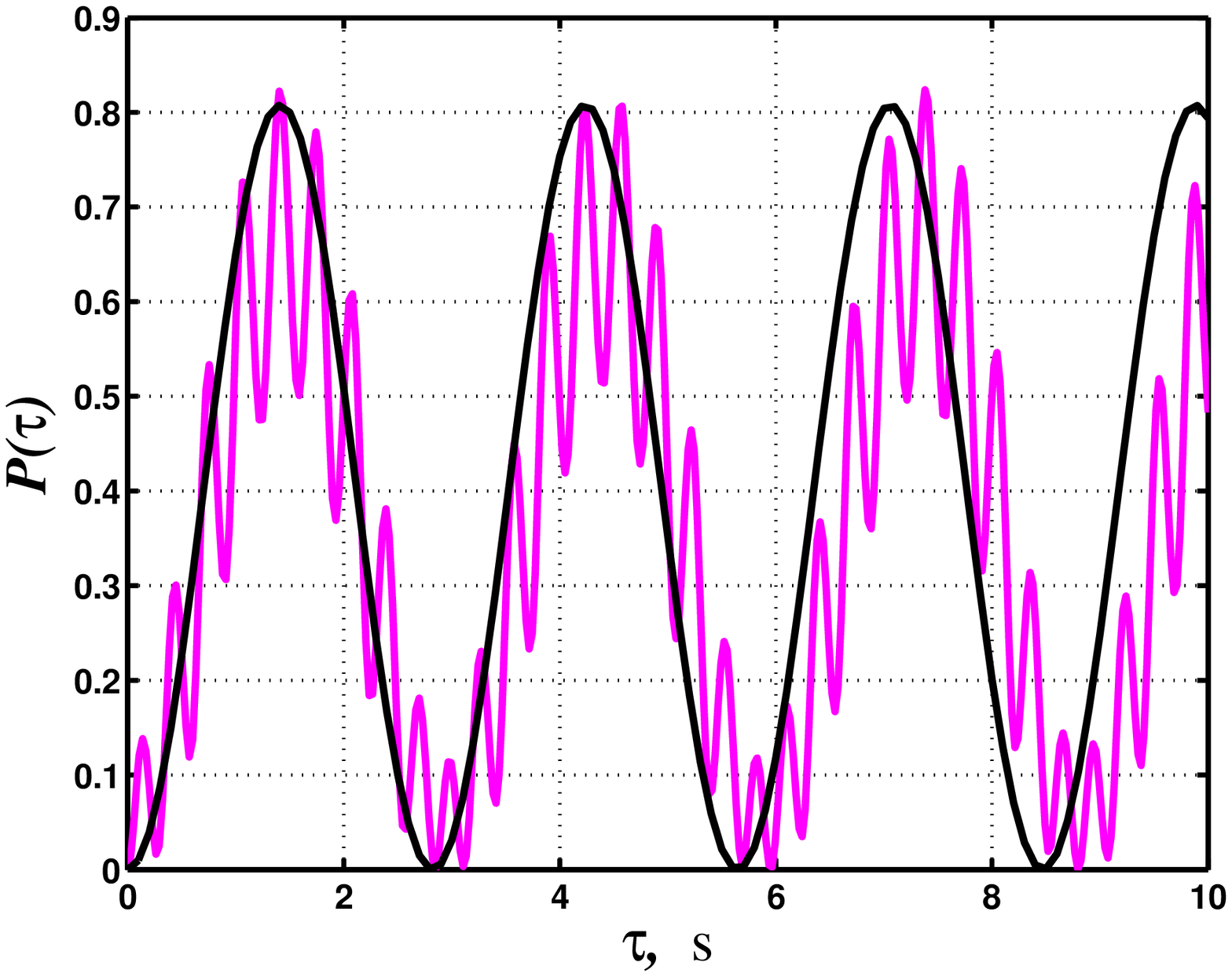}
  \caption{Neutrino transition probability for
  $\nu_{e L}\leftrightarrow\nu_{\mu R}$ oscillations,
  $\omega=2.0\times10^{-14}\thinspace\text{eV}$. The magenta line
  corresponds to the transition probability found with help of
  the numerical solution of Eq.~\eqref{shr}, the black line
  represents the transition probability
  given in Eq.~\eqref{transprob}.}
  \label{numtrpr2}
\end{figure}

\begin{figure}
  \centering
  \includegraphics[scale=.45]{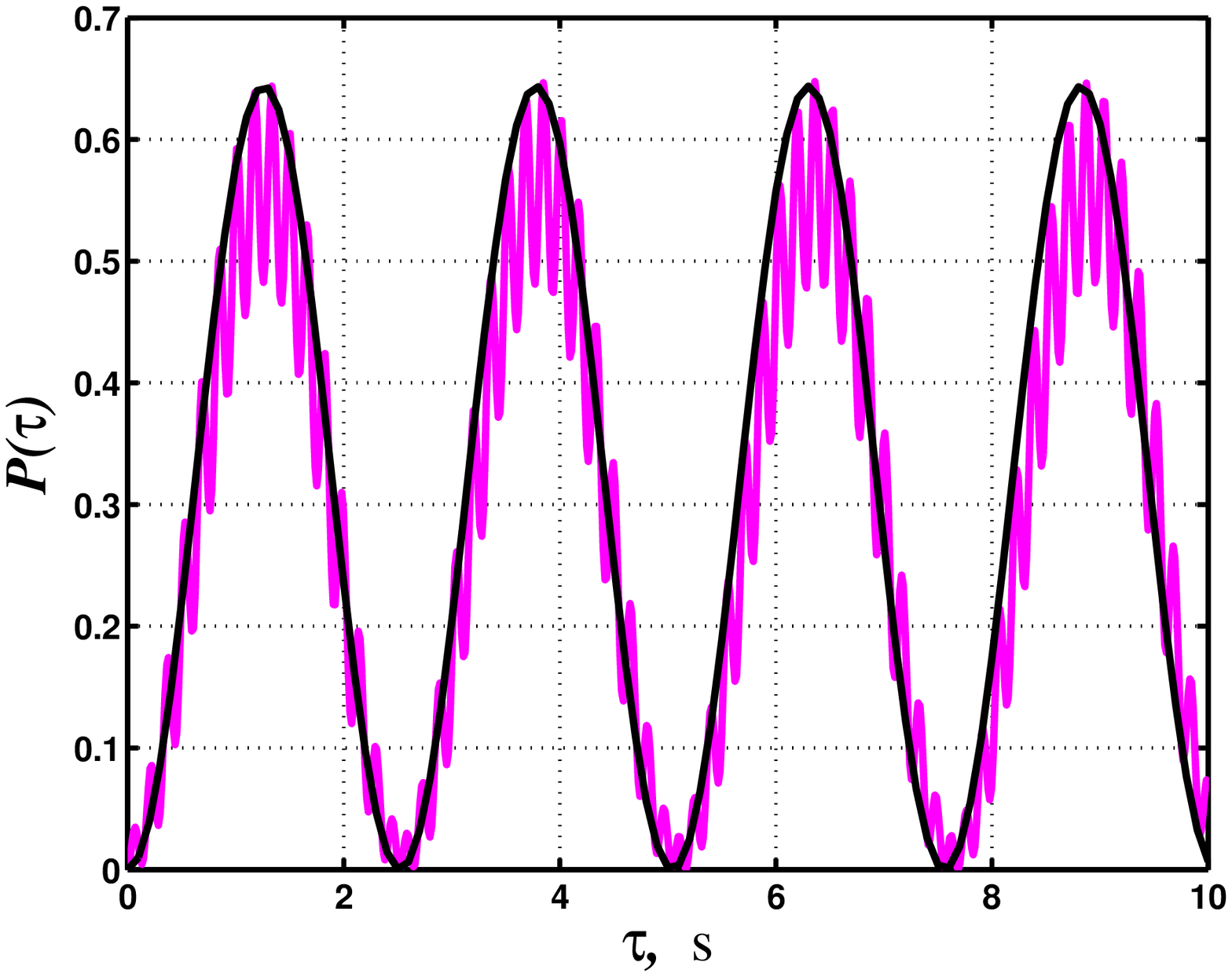}
  \caption{Neutrino transition probability for
  $\nu_{e L}\leftrightarrow\nu_{\mu R}$ oscillations,
  $\omega=4.0\times10^{-14}\thinspace\text{eV}$. The magenta line
  corresponds to the transition probability found with help of
  the numerical solution of Eq.~\eqref{shr}, the black line
  represents the transition probability
  given in Eq.~\eqref{transprob}.}
  \label{numtrpr3}
\end{figure}

The measurable quantity of neutrino oscillations is the transition
probability. The probabilities of the neutrino oscillations are
shown on Figs.~\ref{numtrpr1}-\ref{numtrpr3} versus $\tau=1.5 t$.
We also compare them with the approximate analytical formula
\eqref{transprob}. We remind that we take
$V/2=10^{-15}\thinspace\text{eV}$, $B_0=17.6\thinspace\text{kG}$,
$B=56.3\thinspace\text{kG}$. The resonance frequency is
$\omega=1.0\times 10^{-14}\thinspace\text{eV}$. The magenta line
corresponds to the transition probability computed with help of
the numerical solution of Eq.~\eqref{shr} for the taken
parameters. The black line represents the transition probability
given in Eq.~\eqref{transprob}.

It is noticed in Sec.~\ref{GF} that the small factors
$\mathcal{H}\xi$ and $H_0\xi$ can contribute to the phase of the
neutrino oscillations. However if we consider only one period of
the neutrino transition probability variation, then there is
rather good agreement between numerical and approximate analytical
solutions. Moreover it clearly follows from
Figs.~\ref{numtrpr1}-\ref{numtrpr3} that the phase shift decreases
if $\omega$ becomes greater. Therefore it can be seen in
Figs.~\ref{numtrpr1}-\ref{numtrpr3} that the numerical expression
for the transition probability approaches to the approximate
formula given in Eq.~\eqref{transprob} at great frequencies of the
twisting magnetic field. This comparison proves the validity of
the elaborated technique.

At the end of this section we note that the transition probability
presented in Fig.~\ref{numtrpr1} corresponds to the parameters
that almost coincide with the resonance ones (see
Sec.~\ref{NSFOB}). Despite of the rather low frequency
($\omega=1.0\times10^{-13}\thinspace\text{eV}$), there is a good
agreement between the predicted oscillations parameters such as
amplitude of the transition probability [Eq.~\eqref{A}] as well as
oscillations length [Eq.~\eqref{osclength}] and those obtained
from the numerical solution. The best parameters coincidence
happens within the first period of transition probability
variation.

\section{Conclusion}

In conclusion we note that in this paper we have examined neutrino
oscillations in general rapidly varying fields. We have started
with the usual Schr\"{o}dinger type evolution equation. The
Hamiltonian involved both slowly and rapidly varying in time
terms. Instead of solving the evolution equation directly we have
derived the new effective Hamiltonian which described the
evolution of the averaged neutrino wave function. This new
effective Hamiltonian enabled one to study neutrino oscillations
in arbitrary rapidly varying external fields. It is worth
mentioning that, in contrast to
Refs.~\cite{DvoStu01YFeng,DvoStu04YFeng}, the strength of rapidly
varying fields has not been limited. Therefore the elaborated
method was beyond the perturbation theory since we have not
carried out any expansions over the strength of the external
fields. We have examined the resonance conditions in the case of
general rapidly varying external fields. It has been revealed that
matter with rapidly varying density caused no effect to neutrino
oscillations. We have demonstrated that rapidly varying
axial-vector (e.g., interaction with moving or polarized matter),
tensor and pseudotensor external fields could result in nontrivial
effects in neutrino oscillations. On the basis of the derived new
effective Hamiltonian we have considered neutrino spin-flavor
oscillations in the combination of constant transversal and
twisting magnetic fields. The new effective Hamiltonian for this
system has been obtained. We have analyzed various limiting cases
($B_0\gg B$ and $B\gg B_0$). The limit $B_0\to 0$ has also been
discussed. It has been shown that described mechanism of neutrino
oscillations could be important for the neutrino spin-flavor
conversion in magnetic fields of the Sun. We have evaluated the
strengths of magnetic fields and frequency of the twisting
magnetic field necessary for the $10\%$ neutrino conversion from
$\nu_{e L}$ to $\nu_{\mu R}$. The numerical solutions of the exact
(without neglecting any terms) Schr\"{o}dinger equation for the
two neutrino system interacting with constant transversal and
twisting magnetic fields have been found. We have compared them
with the approximate analytical formula for the transition
probability. The very good agreement at high oscillations
frequencies has been established.

\begin{acknowledgments}
This research was supported by grant of Russian Science Support
Foundation. The author is indebted to Victor Semikoz and all
participants of the IZMIRAN theoretical department seminar for
helpful discussions.
\end{acknowledgments}

\bibliography{generaleng}

\end{document}